\documentstyle[prd,aps,epsf]{revtex}

\draft
\begin{document}

\thispagestyle{empty}
{\baselineskip0pt
\leftline{\large\baselineskip16pt\sl\vbox to0pt{\hbox{\it Department of Physics}
               \hbox{\it Kyoto University}\vss}}
\rightline{\large\baselineskip16pt\rm\vbox to20pt{\hbox{KUNS 1500}
\vss}}%
}
\vskip1cm
\begin{center}{\large \bf
Naked singularity formation in the collapse of a spherical cloud of
counterrotating particles}
\end{center}
\vskip1cm
\begin{center}
 {\large 
Tomohiro Harada 
\footnote{ Electronic address: harada@tap.scphys.kyoto-u.ac.jp},
Hideo Iguchi
\footnote{ Electronic address: iguchi@tap.scphys.kyoto-u.ac.jp},
and Ken-ichi Nakao
\footnote{ Electronic address: nakao@tap.scphys.kyoto-u.ac.jp}
} \\
{\em Department of Physics,~Kyoto University,} 
{\em Kyoto 606-8502,~Japan}\\
\end{center}

\begin{abstract}
We investigate the collapse of a spherical cloud of
counter-rotating 
particles. 
An explicit solution for metric functions
is given using an elliptic integral.
If the specific angular momentum $L(r) =O( r^2)$
at $r\to 0$, 
no central singularity occurs.
With $L(r)$
like that, there is a 
finite region
around the center that bounces.
On the other hand, if the order of $L(r)$ is higher than that,
a central singularity occurs.
In marginally bound collapse with 
$L(r)=4F(r)$, a naked singularity occurs, where $F(r)$ is 
the Misner-Sharp mass.
The solution for this case is expressed 
by elementary functions.
For $ 4 <L/F<\infty$ at $r\to0$,
there is a finite region around the center that bounces and 
a naked singularity occurs. 
For $ 0 \le L/F< 4$ at $r\to0$,
there is no such region.
The results 
suggest that rotation
may play a crucial role on the final fate of collapse.

\end{abstract}
\pacs{PACS numbers: 04.20.Dw, 04.20.Jb, 04.40.-b}

The final fate of gravitational collapse is 
an important problem of relativistic astrophysics and
gravitational physics.
A black hole is usually considered as the final state
of generic gravitational collapse,
and its properties have been investigated~\cite{he1973}.
In particular, a cosmic censorship hypothesis~\cite{penrose1979} is 
a critical assumption of theorems on black holes.
For example, singularity theorems~\cite{he1973}
with this hypothesis
predict the existence of black holes.
Therefore, it is very important to know the behavior of 
collapse of various kinds of matter 
in order to obtain the overall picture about the
general nature of gravitational collapse.
For example, it 
is well known that spherically symmetric dust collapse
from generic initial data
results in a naked 
singularity~\cite{es1979,christodoulou1984,jd1993,sj1996jjs1996}.
This system is rather tractable
because of the existence of an explicit expression, i.e., 
the Lema\^{\i}tre-Tolman-Bondi (LTB) solution.
The assumption of dust matter will be considered to
be rather unrealistic.
In fact, it is obvious that the effect of pressure 
should be taken into account because 
the formation of a naked singularity
in the LTB solution results in blow up of the
energy density.
Introducing tangential pressure alone 
does not lose the merit 
because this system is also given by an explicit
integral~\cite{magli1997b}.
Singh and Witten~\cite{sw1997} considered tangential pressure
proportional to the energy density and found that in collapse
from rest there is a finite region near the center 
which expands outwards
if the tangential pressure is positive. 
A spherical cloud of counter-rotating particles is an example
with tangential pressure but no radial pressure.
This system was considered by Datta~\cite{datta1970}, 
Bondi~\cite{bondi1971} and Evans~\cite{evans1976}.
However, the causal structure of the space-time
was not investigated and this is one of our interests.

Using comoving coordinates, a spherically symmetric space-time 
describing a matter with no radial pressure
is given by
\begin{eqnarray}
  ds^2&=&-e^{2\nu}dt^2+\frac{(R^{\prime})^2 h^2}{1+f}dr^2+R^2(d\theta^2
  +\sin^2\theta d\phi^2), 
\end{eqnarray}
where $h=h(r,R)$ gives a relation between the 
energy density $\epsilon(r,t)\equiv-T^t_t$ and the tangential pressure
$\Pi(r,t)\equiv T^{\theta}_{\theta}=T^{\phi}_{\phi}$ as
\begin{equation}
  \Pi=-\frac{R}{2h}\frac{\partial h}{\partial R}\epsilon,
\end{equation}
and
$\nu=\nu(r,t)$ and $R=R(r,t)$ satisfy the following
set of coupled partially derivative equations: 
\begin{eqnarray}
  \nu^{\prime}&=&-\frac{1}{h}\frac{\partial
    h}{\partial R}
  R^{\prime}.
  \label{eq:comlapse}\\
  \dot{R}^2 
  e^{-2\nu}&=&-1+\frac{2F}{R}+\frac{1+f}{h^2}.
  \label{eq:comenergy}
\end{eqnarray}
The prime and overdot denote the derivative with respect ot $r$ and $t$,
respectively.
The energy density is given by
\begin{equation}
  \epsilon=\frac{F^{\prime}}{4\pi R^2 R^{\prime}}
    \label{eq:density}
\end{equation}
The arbitrary functions $F(r)$ and $f(r)>-1$ are the conserved 
Misner-Sharp mass and the specific energy, respectively.
The function $h(r,R)>0$ has a meaning of the 
internal elastic energy per volume 
and the dust limit is given 
by $h=1$~\cite{magli1997a}.
From regularity at the center,
$  f(0)=h^2(0,0)-1$,
$  R(0,t)=0$,
$  |\nu(0,t)| < \infty$ and
$  F(r)/r^3 < \infty $ at $r\to 0$.
Note that we can set $f(0)=h^2(0,0)-1=0$ because only the
ratio of $h^2$ to $1+f$ is meaningful. 
The solution can be matched with the Schwarzschild space-time
at an arbitrary radius $r=r_b$ if we identify the Schwarzschild
mass parameter $M$ with $F(r_b)$.
Assuming that $R$ is initially a monotonically increasing function 
of $r$ and rescaling the radial coordinate $r$, 
we identify $r$ with the
circumferential radius $R$ on the initial space-like hypersurface
$t=0$.
Here we search the location of an apparent horizon
in a collapsing phase.
Along a future-directed outgoing null geodesic,
\begin{equation}
  \frac{dR}{dt}=\dot{R}
  +R^{\prime}\frac{dr}{dt}
  =e^{\nu}\left[-\sqrt{-1+\frac{2F}{R}+\frac{1+f}{h^2}}
  +\sqrt{\frac{1+f}{h^2}}\right].
\end{equation}
Therefore $R=2F$ is an apparent horizon, $0\le R<2F$ is a trapped region,
and $2F<R$ is an untrapped region.

Tangential pressure for a spherical cloud of 
counter-rotating particles is given 
by~\cite{datta1970,bondi1971,evans1976}
\begin{equation}
  \Pi=\frac{1}{2}\frac{L^2(r)}{R^2+L^2(r)}\epsilon,
\end{equation}
where $L(r)$ is the specific angular momentum.
Then, the function $h$ is given by
\begin{equation}
  h^2=\frac{R^2+L^2}{R^2}.
\end{equation}

Because of the coupling of Eqs. (\ref{eq:comlapse})
and (\ref{eq:comenergy}), the solution given above is not
explicit. Here, according to the procedure of 
Magli~\cite{magli1997b},
we introduce the mass-area coordinates and give an explicit 
form for the general solution.
The space-time is written in these coordinates as
\begin{equation}
  ds^2= -A dm^2- 2B dRdm -C dR^2 +R^2 (d\theta^2 +\sin^2\theta
  d\phi^2),
\end{equation}
where $m$ agrees with the Misner-Sharp mass $F(r)$.
Solving the Einstein equations, we obtain the metric functions
$A(m,R)$, $B(m,R)$ and $C(m,R)$ as follows:
\begin{eqnarray}
  A&=&H(1-\frac{2m}{R}),
  \label{eq:dattabondi1}\\
  B&=&-\frac{ER\sqrt{H}}{|u|\sqrt{R^2+L^2}},\\
  C&=&\frac{1}{u^2}, \\
  u&\equiv&\frac{dR}{d\tau}=\pm\sqrt{-1+\frac{2m}{R}
    +\frac{E^2R^2}{R^2+L^2}},\\
  \sqrt{H(m,R)}&=&\frac{(F^{-1})_{,m}\sqrt{(F^{-1})^2+L^2}}{E F^{-1}}
  \left(\frac{E^2(F^{-1})^2}{(F^{-1})^2+L^2}-1
    +\frac{2m}{F^{-1}}\right)^{-1/2} \nonumber \\
  & &{} +\int^{R}_{F^{-1}}\frac{\sqrt{x^2+L^2}}{x^2 E}
  \left[1+\frac{x^3 E^2}{2(x^2+L^2)}\left(
      \frac{(E^2)_{,m}}{E^2}-\frac{(L^2)_{,m}}{x^2+L^2}\right)\right] 
  \nonumber \\
  & &\cdot\left(-1+\frac{2m}{x}+\frac{E^2x^2}{x^2+L^2}\right)^{-3/2} dx,
  \label{eq:dattabondi4}
\end{eqnarray}
where $\tau$ is a proper time of a comoving observer
at a mass shell labeled by $r$.
The upper and lower signs correspond to expanding and collapsing
phases, respectively.
$F^{-1}$ is the inverse function of $F$
and the existence of it is guaranteed by the 
positivity of the energy density, which hereafter we assume.
We have defined $E^2(m)\equiv 1+f(F^{-1}(m))$.
In order to determine an arbitrary function
which appears in the expression of $\sqrt{H}$,
we have set $R=F^{-1}(m)$ on the initial 
space-like hypersurface, which corresponds to $R(r,0)=r$. 
We have rewritten $L(F^{-1}(m))$ as $L(m)$.
The energy density is given by
\begin{equation}
  \epsilon=\frac{\sqrt{R^2+L^2}}{4\pi |u| R^3 E \sqrt{H}}.
\end{equation}
Then, a shell-crossing singularity occurs when $u\sqrt{H}=0$.
If a shell-crossing occurs, the coordinate system breaks down
after that. 
The integral in the expression of $\sqrt{H}$ is reduced to
\begin{equation}
  \int^{R}_{F^{-1}}\frac{2(x^2+L^2)^2+x^3[(E^2)_{,m}
      (x^2+L^2)-(L^2)_{,m}E^2]}{2E[(E^2-1)x^3+2mx^2-L^2x+2mL^2]
      \sqrt{(E^2-1)x^4+2mx^3-L^2x^2+2mL^2x}} dx.
\end{equation}
Therefore we find that the general solution
is expressed by an elliptic integral.

If $L(m)=0$, the solution is reduced to the 
LTB solution in the mass-area coordinates.
If $E(m)=1$ (which is called marginally bound collapse)
and $L(m)=4m$, 
the integral is expressed by elementary functions as follows:
\begin{eqnarray}
  A&=&H(1-\frac{2m}{R}),
  \label{eq:l4f1}\\
  B&=&-\frac{R}{|R-4m|}\sqrt{\frac{RH}{2m}},\\
  C&=&\frac{R(R^2+16m^2)}{2m(R-4m)^2}, \\
  u&=&\pm|R-4m|\sqrt{\frac{2m}{R(R^2+16m^2)}},\\
  \sqrt{H}&=&\frac{(F^{-1})_{,m}((F^{-1})^2+16m^2)}
  {|F^{-1}-4m|\sqrt{2mF^{-1}}} \nonumber \\
  & &{} +\mbox{sign}(F^{-1}-4m)\left[\left(
      \frac{R^2-16 m R+144m^2}{3\sqrt{2}m(R-4m)}\sqrt{\frac{R}{m}}
      +4\sqrt{2}\ln\frac{\sqrt{R}+2\sqrt{m}}
      {|\sqrt{R}-2\sqrt{m}|}
      \right)\right. \nonumber \\
  & &\left.{}-\left(
      \frac{(F^{-1})^2-16 m F^{-1}+144m^2}{3\sqrt{2}m(F^{-1}-4m)}
      \sqrt{\frac{F^{-1}}{m}}
      +4\sqrt{2}\ln\frac{\sqrt{F^{-1}}+2\sqrt{m}}
      {|\sqrt{F^{-1}}-2\sqrt{m}|}
    \right)
  \right].
  \label{eq:l4f4}
\end{eqnarray}

Now that we have obtained the explicit expression for the metric
functions, we investigate the singularity that may occur in the collapse
of a spherical cloud of counter-rotating particles.
Since shell-crossing singularities are considered to be 
gravitationally weak,
we concentrate on shell-focusing singularities
which is defined by $R=0$.
Furthermore, no light ray can emmanate from
non-central ($r>0$) shell-focusing
singularities because $R=0<2F$ 
for $r>0$ and hence it is covered~\cite{christodoulou1984}.
From the above discussion,
it turns out to be sufficient 
to consider a central ($r=0$) shell-focusing singularity
in order to discuss whether or not a naked singularity exists.
The motion of each shell of $r>0$ 
is described by Eq. (\ref{eq:comenergy}),
i.e.,
\begin{equation}
  \left(\frac{dR}{d\tau}\right)^2=-1+\frac{2F}{R}+\frac{(1+f)R^2}{R^2+L^2}.
  \label{eq:eom}
\end{equation}
By investigating the shape of the effective potential 
\begin{equation}
  V(R;r)\equiv-\frac{2F}{R}+1-\frac{(1+f)R^2}{R^2+L^2},
  \label{eq:effpot}
\end{equation}
we can get qualitative understanding about 
the motion of each shell of $r>0$.

Proceeding to deal with the symmetric center $r=0$,
we take the analyticity there into consideration.
We assume 
that the metric functions and the function $h^2(r,R)$
are $C^{\infty}$ class at least in the neighborhood of the 
center $r=0$
before encountering a central singularity. 
From this assumption, the metric variables
in the comoving coordinates are expanded as
\begin{eqnarray}
  \nu(r,t)&=&\nu_0(t)+\nu_2(t) r^2+\nu_4(t)r^4+\cdots, \\
  R(r,t)&=&R_1(t)r+R_3(t)r^3+R_5(t)r^5+\cdots,
\end{eqnarray}
and we can set $\nu_0(t)=0$ by using the rescaling
freedom of the time coordinate.
Then, from Eqs. (\ref{eq:comlapse}) and (\ref{eq:comenergy}),
the arbitrary functions $F(r)$, $f(r)$ and $L^2(r)$ 
should be expanded as
\begin{eqnarray}
  F(r)&=&F_3r^3+F_5r^5+\cdots, \\
  f(r)&=&f_2r^2+f_4r^4+\cdots, \\
  L^2(r)&=&L_4r^4+L_6r^6+\cdots,
\end{eqnarray}
and, from Eq. (\ref{eq:density}), 
the energy density should be expanded as
\begin{equation}
  \epsilon(r,t)=\epsilon_0(t)+\epsilon_2(t)r^2+\epsilon_4(t)r^4+\cdots,
\end{equation}
and then
\begin{equation}
  \epsilon(0,t)=\epsilon_0(t)=\frac{3F_3}{4\pi R_1(t)^3}.
\end{equation}
Observing the lowest order of Eq. (\ref{eq:comenergy}), 
the time development of $R_1(t)$ is given by
\begin{equation}
  \left(\frac{dR_1}{dt}\right)^2=\frac{2F_3}{R_1}+f_2-\frac{L_4}{R_1^2}.
  \label{eq:center}
\end{equation}  
Since $L^2(r)\ge 0$, $L_4 \ge0$.

First we consider the case of $L_4>0$. 
From Eq. (\ref{eq:center}) we find that
$R_1(t)$ cannot vanish.
This result was firstly shown by Evans~\cite{evans1976}.
Therefore no central singularity occurs
for $L_4>0$.
Next we examine the motion of the shell $r>0$.
From Eq. (\ref{eq:effpot}), allowed regions for a given
$r$ are obtained by
\begin{equation}
  g(R;r)\equiv -f R^3 -2F R^2 +L^2 R -2 F L^2 \le 0.
\end{equation}
For sufficiently small $r>0$, 
\begin{eqnarray}
  g(R=0;r)\approx -2FL^2 <0, \\
  g(R=4F;r)\approx 2FL^2 >0,
\end{eqnarray}
where ``$\approx$'' means the equality up to the lowest order.
Considering that $g(R;r)$ is a cubic function of $R$,
the allowed regions are given by
\begin{eqnarray}
  0\le R\le R_{1},~~R_{2}\le R,~~(\mbox{for}~~f\ge 0),\\
  0\le R\le R_{1},~~R_{2}\le R\le R_{3},~~(\mbox{for}~~f < 0),
\end{eqnarray}
where the following inequality is satisfied
\begin{equation}
  0<R_{1}<4F<R_{2}.
\end{equation}
Since $F=O(r^3)$, $R(r,t=0)=r$ cannot be in the inner allowed 
region $0\le R \le R_{1}$.
This means that $R(r,t)$ must be in 
the outer allowed region at $t=0$.
Therefore we conclude that the region around $r=0$, 
which was initially in a collapsing phase, 
necessarily experiences a bounce and begins to expand.
The motion after that is an eternal expansion for $f\ge 0$
or oscillations for $f<0$.
This implies that, since 
$R>R_{2}>4F >2F$,
the region around $r=0$ is untrapped.
We summarize this case by no central singularity,
no apparent horizon and
bounce of the region around the center.

We proceed to the case of $L_4=0$.
From Eq. (\ref{eq:center}), we find that
the initially collapsing cloud inevitably form
a central shell-focusing singularity after a
finite proper time.
In order to see whether this central singularity
is naked or covered, 
we examine the motion of the region around the center.
Here we define
\begin{equation}
  D\equiv\lim_{r\to 0}\frac{L}{F}=L_6^{1/2}/F_3.
\end{equation}

For $D>4$, 
for sufficiently small $r>0$,
\begin{eqnarray}
  g(R=0;r)\approx -2FL^2 <0, \\
  g(R=\frac{D^2}{4}F;r)\approx \frac{D^2(D^2-16)}{8}F^3 >0.
\end{eqnarray}
Then,
the allowed regions are
\begin{eqnarray}
0 \le R \le R_{1},~~R_{2} \le R~~(\mbox{for}~~f\ge 0),\\
0 \le R \le R_{1},~~R_{2} \le R \le R_{3}~~(\mbox{for}~~f<0),
\end{eqnarray}
where the following inequality is satisfied:
\begin{equation}
  0<R_{1}<\frac{D^2}{4}F<R_{2}.
\end{equation}
In the same way as for the case of $L_4>0$, 
we conclude that the region around $r=0$, 
which was initially in a collapsing phase, 
necessarily experiences a bounce and begins to expand.
The motion after that is an eternal expansion for $f\ge 0$
or oscillations for $f<0$.
This implies that, since 
$R>R_{2}>4F >2F$,
the region around $r=0$ is untrapped.
Therefore the central singularity is naked.
Since the space-time can be matched to the Schwarzschild space-time
at an arbitrary radius $r=r_b$, we can construct the space-time
with a globally naked singularity.
The results given above does not depend on 
details of initial density distribution.

For $D=4$, the behavior depends on the higher order terms.
Here we present the critical and most interesting 
case in which $f(r)=0$ and $L(r)=4F(r)$.
For this case,
the metric functions are exactly solved and expressed by 
elementary functions as seen in Eqs. (\ref{eq:l4f1})-
(\ref{eq:l4f4}).
In this case the effective potential
is given by
\begin{equation}
  V=-\frac{2F(R-4F)^2}{R(16F^2+R^2)}.
\end{equation}
On the initial space-like hypersurface, regularity requires 
$ R(r,t=0)=r>4F =O( r^3) $
in a sufficiently small but finite region around $r=0$.
Then each initially collapsing shell of $r>0$ 
approaches $R=4F$.
From Eq. (\ref{eq:eom}), the behavior of this approach
is as 
\begin{equation}
  R-4F \propto \exp(-\frac{\tau}{8F}).
\end{equation}
This behavior means that $R$ approaches $4F$ 
asymptotically.
This behavior is also the case in an expanding phase 
for $R<4F$, and therefore the sign of $R-4F$ 
does not change.
That is why
we have chosen the sign as seen in 
Eqs. (\ref{eq:l4f1})-(\ref{eq:l4f4}).
Then, in the region around the center $r=0$,
the collapse suffers the continuing 
angular momentum braking.
Since $R>4F>2F$, the region around the center 
is untrapped eternally.
Therefore the central singularity is naked
and can be globally naked.
Furthermore we prove that a shell crossing does not
occur and the coordinate system is valid at least in the region
around the center by showing that $\sqrt{H}>0$.
Each term in the first parenthesis in the bracket on the right hand side
of Eq. (\ref{eq:l4f4})
is positive, because $R>4F$ in the region around the center.
The contribution of the first term and the terms in the second parenthesis
is expanded as
\begin{equation}
  \frac{\sqrt{2}}{9}\frac{24F_3^2-F_5}{F_3^{13/6}}m^{-1/3} +O(m^{1/3}),
\end{equation}
where terms which is $O( m^{-1})$ cancel each other.
Therefore, if $F_5 < 24 F_3^2$, 
at least a sufficiently small region around the center 
is shell-crossing free, and we can construct a shell-crossing
free solution by matching. 
This condition holds if $\epsilon(r,t=0)$
is decreasing function of $r$.
The nakedness of the central singularity is confirmed by
examining the algebraic 
root equation given by Magli~\cite{magli1997b}.
If there is a finite positive root $x_0$ for this equation
for some $\alpha>1/3$,
the central singularity is naked.
Using the exact solution (\ref{eq:l4f4}), we obtain
a finite positive root
\begin{equation}
  x_0=\left(\frac{24F_3^2-F_5}{4\sqrt{2} F_3^{13/6}}\right)^{2/3},
\end{equation}
for $\alpha=7/9$.
Then, the future-directed outgoing null geodesic starting
from the central singularity behaves as
\begin{equation}
  R\approx 2 x_0 m^{7/9} \approx 2 x_0 F_3^{7/9} r^{7/3}.
\end{equation}
  
For the case $0\le D<4$, the collapse continues to a covered
singularity for $r>0$.
In order to see whether the central singularity is naked or not,
we have to examine the existence of positive and real roots 
of the root equation 
given in the mass-area coordinates by Magli~\cite{magli1997b}
using the explicit form of the 
solution (\ref{eq:dattabondi1})-(\ref{eq:dattabondi4}).
Works for $0\le D<4 $ 
are now in progress.

In summary, we have investigated the final fate of the collapse of 
a spherical cloud of counter-rotating particles.
We have presented an explicit expression 
for the metric functions
and shown that this is given by an elliptic integral.
For marginally bound collapse with angular momentum 
distribution $L(r)=4F(r)$, 
we have succeeded to give an expression by elementary functions.
The existence of the lowest order term in $L(r)$ 
that is allowed by regularity always 
guarantees regularity of the evolved center.
On the other hand, we have shown that the absence of this term
inevitably results in a central singularity in the collapse
of the cloud.
In the former case, the region around the center
necessarily bounces to an expanding phase.
In the latter case, if 
\begin{equation}
\lim_{r\to0}\frac{L}{F} > 4,
\label{eq:lflimit}
\end{equation}
the region around the center $r=0$ bounces to an expanding phase
and the central singularity is 
naked and can be globally naked by matching 
to the Schwarzschild space-time.
If $f(r)=0$ and $L(r)=4F(r)$, the region around the center slows down 
and approaches asymptotically 
to $R=4F$ and the central singularity is also naked.
If the limit in Eq. (\ref{eq:lflimit}) is smaller than 4,
the region around the center collapses to a space-like 
singularity.
These results shows that {\it tangential pressure may undress
the covered singularity}.
In particular, rotation may induce 
the naked singularity
formation.
For example, a sufficiently small, marginally bound, 
and spherical ball of counter-rotating 
particles with $\epsilon(r,t=0)=\mbox{const}$ 
and $L(r)=4F(r)$ 
collapses to a naked singularity.
Our results suggest that the effect of pressure
is not negligible.
Furthermore anisotropic velocity dispersion of gravitating particles
and/or anisotropic pressure which may be realized by crystallization
in a dense nuclear matter may play an important role 
in the final stage of collapse.
 
\acknowledgements
We are grateful to T. Nakamura
and D. Ida for helpful discussions.
We are also grateful to H. Sato
for his continuous encouragement.
This work was partly supported by the 
Grant-in-Aid 
for Scientific Research Fellowship (No.9204) 
and for Creative Basic Research (No.09NP0801)
from the Japanese Ministry of Education, Science, Sports  
and Culture.

\appendix

\end{document}